\documentclass[apjl]{emulateapj}
\usepackage{apjfonts}

\newcommand{\nuclei}[2]{\ensuremath{^{#1}\mathrm{#2}}}
\newcommand{\N}{\nuclei{14}{N}}
\newcommand{\Oxy}{\nuclei{18}{O}}
\newcommand{\F}{\nuclei{18}{F}}
\newcommand{\Ne}{\nuclei{22}{Ne}}
\newcommand{\Fe}[1][54]{\nuclei{#1}{Fe}}
\newcommand{\Ni}[1][56]{\nuclei{#1}{Ni}}
\newcommand{\Ye}{\ensuremath{Y_e}}
\newcommand{\Zsun}{\ensuremath{Z_\odot}}
\newcommand{\SNeIa}{SNe~Ia}
\newcommand{\SNeII}{SNe~II}
\newcommand{\FeH}{\ensuremath{[\mathrm{Fe/H}]}}
\newcommand{\XFe}{\ensuremath{[\mathrm{X/Fe}]}}
\newcommand{\CFe}{\ensuremath{[\mathrm{C/Fe}]}}
\newcommand{\NFe}{\ensuremath{[\mathrm{N/Fe}]}}
\newcommand{\OFe}{\ensuremath{[\mathrm{O/Fe}]}}
\newcommand{\FeFe}{\ensuremath{\Fe/\Fe[56]}}
\newcommand{\ee}[1]{\ensuremath{\times 10^{#1}}}

\newcommand{\unitspace}{\ensuremath{\,}}
\newcommand{\usp}{\unitspace}
\newcommand{\unitstyle}[1]{\ensuremath{\mathrm{#1}}}
\newcommand{\power}[2]{\ensuremath{{#1}^{#2}}}

\newcommand{\Giga}{\unitstyle{G}}
\newcommand{\cm}{\unitstyle{cm}}
\newcommand{\gram}{\unitstyle{g}}
\newcommand{\K}{\unitstyle{K}}	
\newcommand{\grampercc}{\gram\usp\power{\cm}{-3}} 
\newcommand{\GramPerCc}{\grampercc}
\newcommand{\Msun}{\ensuremath{M_\odot}}

\newcommand{\Gyr}{\Giga\yr}
\newcommand{\yr}{\unitstyle{yr}}	
\newcommand{\unit}[2]{\ensuremath{#1\usp\mathrm{#2}}}

\begin{document}

\slugcomment{To appear in \textsc{The Astrophysical Journal Letters}}

%
\title{On Variations in the Peak Luminosity of Type Ia Supernovae}

\author{
F.~X.~Timmes\altaffilmark{1},
Edward F.~Brown\altaffilmark{2}, 
J.~W.~Truran\altaffilmark{1,2} 
} 
\affil{Center for Astrophysical Thermonuclear Flashes, 
       The University of Chicago, 
       Chicago, IL  60637}

\email{fxt@flash.uchicago.edu}

\altaffiltext{1}{Dept.\ of Astronomy \& Astrophysics, 
                 The University of Chicago, 
                 Chicago, IL  60637}

\altaffiltext{2}{Enrico Fermi Institute,
                 The University of Chicago,
                 Chicago, IL 60637}

\begin{abstract}

We explore the idea that the observed variations in the peak
luminosities of Type Ia supernovae originate in part from a scatter in
metallicity of the main-sequence stars that become white dwarfs.
Previous, numerical, studies have not self-consistently explored
metallicities greater than solar.  One-dimensional Chandrasekhar mass
models of \SNeIa\ produce most of their \Ni\ in a burn to nuclear
statistical equilibrium between the mass shells 0.2\usp\Msun\ and
0.8\usp\Msun, for which the electron to nucleon ratio \Ye\ is constant
during the burn.  We show analytically that, under these conditions,
charge and mass conservation constrain the mass of \Ni\ produced to
depend \emph{linearly} on the original metallicity of the white dwarf
progenitor.  Detailed post-processing of W7-like models confirms this
linear dependence.  The effect that we have identified is most evident
at metallicities larger than solar, and is in agreement with previous
self-consistent calculations over the metallicity range common to both
calculations.  The observed scatter in the metallicity
(\onethird\Zsun--3\Zsun) of the solar neighborhood is enough to induce a
25\% variation in the mass of \Ni\ ejected by Type Ia supernova.  This
is sufficient to vary the peak $V$-band brightness by $|\Delta M_V|
\approx 0.2$.  This scatter in metallicity is present out to the
limiting redshifts of current observations ($z \lesssim 1$).
Sedimentation of \Ne\ can possibly amplify the variation in \Ni\ mass to
$\lesssim 50\%$.  Further numerical studies can determine if other
metallicity-induced effects, such as a change in the mass of the
\Ni-producing region, offset or enhance the variation we identify.

\end{abstract}


%
\section{Introduction }

The maximum luminosity of Type Ia supernovae (\SNeIa) is a key
ingredient of their analysis and is also essential for their use as
distance indicators in cosmology (\citealt{filippenko97};
\citealt{branch98}; \citealt{leibundgut01}).  For the nearby
\SNeIa\ with Cepheid-determined distances, the overall dispersion in the
peak magnitude measurements is rather small, about 0.5 mag.\ in $B$ and
$V$ (\citealt{filippenko97}; \citet{saha99}; \citealt{gibson00}).  When
the sample is enlarged to include more distant \SNeIa, there are several
subluminous events that broaden the variation to about 1~mag.\ in $B$
(\citealt{hamuy96}; \citealt{riess98}; \citealt{perlmutter99};
\citealt{phillips99}), but the bulk of the \SNeIa\ sample have peak
brightnesses within a 0.5~mag.\ range in $B$ and $V$.

An interesting feature of \SNeIa\ is that the luminosity is set not by
the explosion, for which the deposited energy goes into expansion, but
rather by the decay of \Ni\ and \nuclei{56}{Co} that are formed during
the nucleosynthesis \citep{arnett82,pinto00}.  At the time of peak
luminosity, \nuclei{56}{Co} has not yet decayed, and hence the peak
luminosity is a measurement of the amount of \Ni\ synthesized during the
explosion.  This amount presumably depends on the progenitor and on the
details of the explosion.  In particular, it has long been known that
the amount of \Ni\ synthesized depends in part on the asymmetry between
neutrons and protons in the progenitor white dwarf (\citealt*{truran67};
\citealt*{arnett71}).

In this Letter we explore how the intrinsic variation in the initial CNO
abundances of the progenitors translates into a variation in the mass of
\Ni\ synthesized ($M(\Ni)$) and hence in the peak luminosity of \SNeIa.
We show analytically, using mass and charge conservation, that $M(\Ni)$
depends \emph{linearly} on the initial metallicity $Z$, and that the
observed variation in the $Z$ of field dwarfs, from \onethird\ to 3
times solar, leads to a $\approx 25\%$ variation in $M(\Ni)$, with most
of the effect occurring at $Z > \Zsun$.  This is the expected variation
if \SNeIa\ progenitors uniformly represent the range of CNO abundances
observed in stars today.  Our conclusion rests on (1) considerations of
nuclear equilibrium during the explosion, and (2) the observed scatter
in $Z$ of stars within the Galactic disk.

The dependence of the ejected \Ni\ mass on the progenitor's initial $Z$
has been investigated previously, both via the evolution of the
progenitor (\citealt{umeda99a}; \citealt{umeda99b};
\citealt*{dominguez01}, hereafter D01) and the explosion itself
(\citealt*{hoeflich98}; \citealt{iwamoto99}; D01).  Although these
one-dimensional simulations are sophisticated in their treatment of the
flame microphysics and the nuclear burning, we desire to elucidate
features that are robust to any complicated hydrodynamics.  Our
demonstration that the mass of nickel depends linearly on the initial
$Z$ of the progenitor serves as a check and stimulus for future
numerical studies and provides insight into possible evolutionary
effects.

In \S~\ref{sec:variation-Ni-mass} we demonstrate the linear dependence
of $M(\Ni)$ on the progenitor's $Z$, and compare this with detailed
nucleosynthesis calculations and previous numerical work.  We speculate
on the source of differences between different calculations, and then
discuss the intrinsic scatter in CNO abundances in the ISM
(\S~\ref{sec:scatter-metallicity}) and the implications
(\S~\ref{sec:implications}) of our results.

\section{Variations in the Mass of Nickel Ejected}
\label{sec:variation-Ni-mass} 

Nearly all one-dimensional Chandrasekhar mass models of \SNeIa\ produce
most of their \Ni\ in a burn to nuclear statistical equilibrium (NSE)
between the mass shells 0.2\usp\Msun\ and 0.8\usp\Msun\
(\citealt*{nomoto84}; \citealt{hoeflich98}; \citealt{iwamoto99};
\citealt{hoeflich00}).  In this region, unlike in the innermost
0.2\usp\Msun\ \citep{brachwitz00}, weak interactions operate on
timescales longer than the time for the thermonuclear burning front to
disrupt the white dwarf.  Following this rapid burn to NSE, most of the
mass is in the iron-peak nuclei \Ni, \Ni[58], and \Fe.  First consider
the case when \Ni\ and \Ni[58] are the only two competing species.  The
addition of \Fe\ will be considered next.  Mass and charge conservation,
\begin{equation}
 \sum_{i=1}^n X_i = 1,
 \qquad
 \sum_{i=1}^n \frac{Z_i}{A_i} X_i = \Ye
 \label{eq:mqconserv}
\end{equation}
imply that the mass fraction of \Ni\ depends linearly on \Ye,
\begin{equation}
 X(\Ni) = 58 \Ye - 28 ,
 \label{eq:nse1}
\end{equation}
where isotope $i$ has $Z_i$ protons, $A_i$ nucleons (protons +
neutrons), and a mass fraction $X_i$. The aggregate ensemble has a
proton to nucleon ratio of \Ye.

Most of a main-sequence star's initial $Z$ comes from the CNO
and \Fe[56] nuclei inherited from its ambient interstellar medium. The
slowest step in the hydrogen burning CNO cycle is proton capture onto
\N. This results in all the CNO catalysts piling up into \N\ when
hydrogen burning on the main sequence is completed. During helium
burning the reactions
$\N(\alpha,\gamma)\F(\beta^{+},\nu_e)\Oxy(\alpha,\gamma)\Ne$ convert all
of the \N\ into \Ne. Thus, the mass fraction of \Ne\ in the
carbon-oxygen white dwarf remnant is
\begin{equation}
 X(\Ne) = 22 \left( 
                    \frac{X(\nuclei{12}{C})}{12} 
                  + \frac{X(\N)}{14} 
                  + \frac{X(\nuclei{16}{O})}{16} 
                        \right) ,
 \label{eq:x22}
\end{equation}
where the mass fractions refer to the original distribution of the star
(prior to main-sequence burning).  For a uniform distribution of \Ne\
and \Fe[56] throughout the star, eq.~(\ref{eq:mqconserv}) gives the
initial \Ye\ of the white dwarf
\begin{equation}
Y_e  =  \frac{10}{22} \, X(\Ne) + \frac{26}{56} \, X(\Fe[56]) 
      + \frac{1}{2} \, \left[ 1 - X(\Ne) - X(\Fe[56]) \right] .
 \label{eq:ye_initial}
\end{equation}
According to the argument at the beginning of this section, this \Ye\ is
fixed in the region where the \Ni\ is created.  Substituting
eqs.~(\ref{eq:ye_initial}) and (\ref{eq:x22}) into eq.~(\ref{eq:nse1})
leads to a linear expression for the mass fraction of \Ni\ in an NSE
distribution in terms of the main-sequence star's initial $Z$
\begin{eqnarray}
  X(\Ni) &=& 1 - 58 \left[\frac{X(\nuclei{12}{C})}{12} +
    \frac{X(\N)}{14} + \frac{X(\nuclei{16}{O})}{16} +
    \frac{X(\Fe[56])}{28} \right]
      \nonumber \\
           & = & 1 - 0.057 \frac{Z}{\Zsun}.
 \label{eq:x56}
\end{eqnarray}
The average peak $B$ and $V$ magnitudes of nearby \SNeIa\
\citep{saha99,gibson00} strongly imply that a fiducial \SNeIa\ produces
$\approx 0.6\usp\Msun$ of \Ni.  Taking eq.~(\ref{eq:x56}) to
represent the mass fraction of \Ni\ relative to this fiducial mass gives
\begin{equation}
  M(\Ni) \approx 0.6\usp\Msun \left( 1 - 0.057\frac{Z}{\Zsun} \right).
 \label{eq:m56}
\end{equation}
Here we assume that \Ye\ is uniform throughout the star, and that all
material within the \Ni-producing mass shell passes through NSE with
normal freeze-out.

If \Fe\ is also present then an additional Saha-like equation for the
chemical potentials \citep{clifford65} is required. Consider the
chemical equation $ \alpha\Ni + \beta\Fe \leftrightarrow
\gamma\nuclei{58}{Ni} $.  Both \Fe\ and \nuclei{58}{Ni} carry two extra
neutrons; this fixes the stoichiometric ratios $\gamma/\beta = 1$ and
$\beta/\alpha = 14$.  As a result, the equation for chemical equilibrium
is of the form $ X(\Fe)/X(\nuclei{58}{Ni}) = [f(T,\rho)/X(\Ni)]^{1/14}
$.  Here $f(T,\rho)$ contains the mass excesses and phase space factors
common to the chemical potential of the three species.  Expanding about
$T = 5\ee{9}\usp\K$, $\rho = 10^8\usp\GramPerCc$, and $X(\Ni) = 1.0$, we
find that
\begin{eqnarray}\label{eq:Fe54-Ni58}
   \frac{X(\Fe)}{X(\nuclei{58}{Ni})} &\approx& 2.1
   \left( \frac{T}{5\ee{9}\usp\K} \right)^{0.4}
   \left( \frac{\rho}{10^8\usp\GramPerCc} \right)^{-0.07} \nonumber \\
   && \times X(\Ni)^{-0.07}.
\end{eqnarray}
For this ratio, $X(\Fe)/X(\Ni[58]) \approx 2$, the relationship between
$Z$ and $X(\Ni)$ becomes slightly shallower than eq.~(\ref{eq:x56}),
$X(\Ni) = 1 - 0.054 (Z/\Zsun)$.  Similar considerations show that
\Fe[56] is not present unless \Ye\ is much smaller than the case
considered.  Equation~(\ref{eq:Fe54-Ni58}) does not hold if an
alpha-rich freeze-out occurs, in which case $(\alpha,\gamma)$ reactions
convert \Fe\ to \nuclei{58}{Ni}.  Hence the ratio $X(\Ni[58])/X(\Fe)$ is
sensitive to the flame speed.  The \Ni\ mass is affected only slightly,
however, as the charge-to-mass of \nuclei{58}{Ni} (which sets the slope
in eq.~[\ref{eq:nse1}]) is $28/58 = 0.483$ and differs by only 0.3\%
from the charge-to-mass of \Fe, $26/54 = 0.481$.  Our main result,
eq.~(\ref{eq:x56}), is thus relatively insensitive to the precise
details of the explosion, so long as NSE favors a \Ni-dominated ``iron''
peak.

\begin{figure}[tb]
\plotone{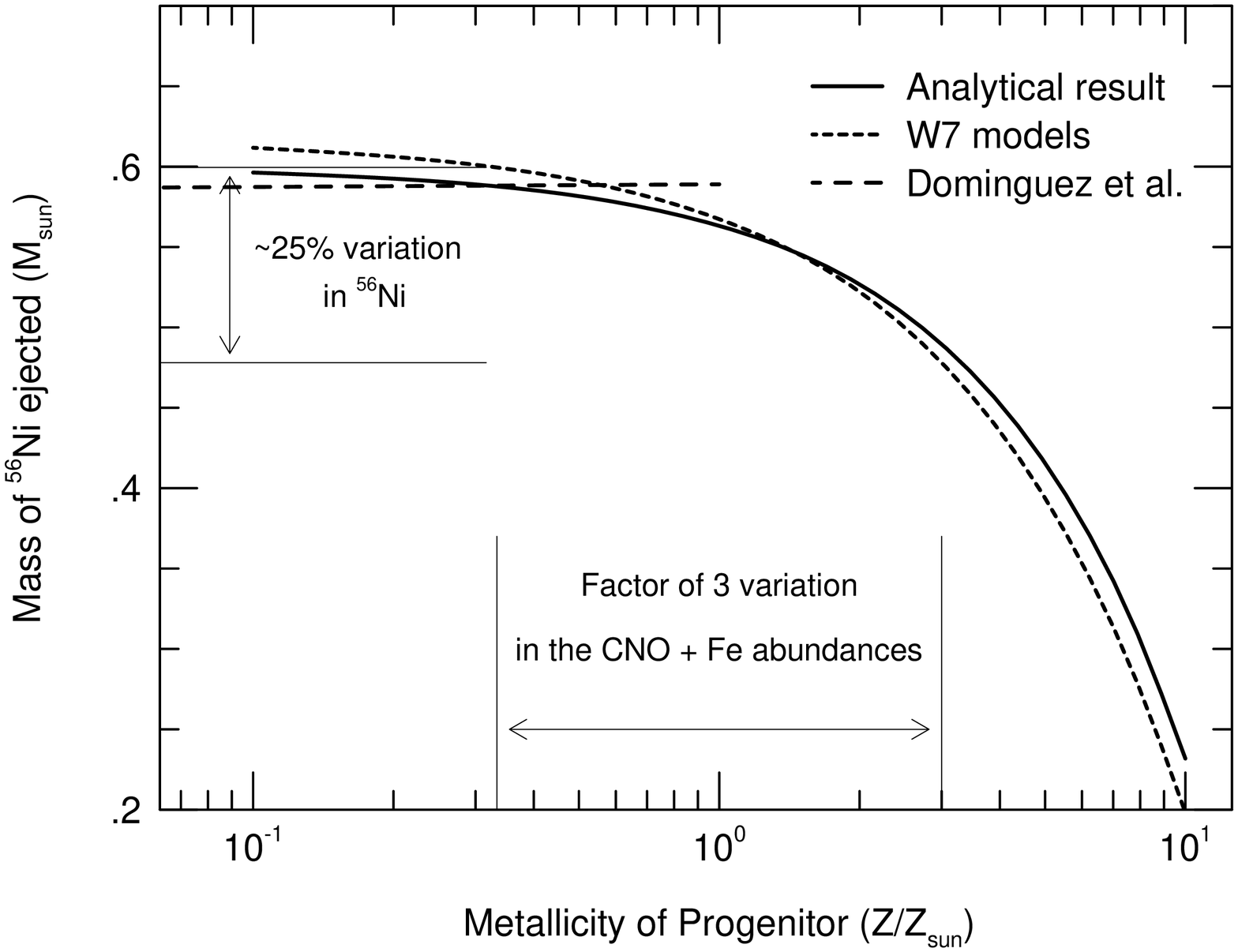}
\caption{\label{fig:mni56}%
  Mass of \Ni\ ejected by \SNeIa\ as a function of the initial
  metallicity $Z$.  Shown is the linear relation (\emph{solid curve};
  the curvature is from the logarithmic abscissa) of
  eq.~(\ref{eq:m56}) for $X(\Ne) = 0.024(Z/\Zsun)$, a sequence of
  W7-like models (\emph{short-dashed curve}), and the calculation of D01
  for 1.5\usp\Msun progenitors (\emph{long-dashed curve}).  Other
  progenitor masses in the D01 survey display the same trend with
  $Z$.  As indicated by the arrows, a scatter of 3 about the
  mean in $Z$ of the main-sequence stars that produce white dwarfs leads
  to a variation of about 25\% (0.13\usp\Msun) of \Ni\ ejected if the
  metals are uniformly distributed within the white dwarf.  A factor of
  seven scatter about the mean in the initial metallicity corresponds to
  a factor of 2 variation in $M(\Ni)$.  }
\end{figure}

The simple linear relation between \Ye\ and the mass fraction of \Ni\ in
eq.~(\ref{eq:x56}) is robust, as it only relies on basic properties
of NSE.  As a test of this analytical result, we calculated the mass of
\Ni\ ejected by W7-like models (\citealt{nomoto84};
\citealt*{thielemann86}; \citealt{iwamoto99}) by integrating a 510
isotope nuclear reaction network over the thermodynamical trajectories
for a wide range of $Z$.  The results are shown in Fig.~\ref{fig:mni56}
with the short-dashed curve.  We also plot the linear relation
(eq.~[\ref{eq:m56}]; \emph{solid curve}), with $Z/\Zsun$ adjusted to
give the \Ne\ abundance used by W7, $X(\Ne) = 0.025\,Z/\Zsun$.  Most of
the differences between eq.~(\ref{eq:m56}) and the detailed W7-like
models is attributable to our assumption that all of the \Ni\ comes from
the 0.2 to 0.8\usp\Msun\ region in the white dwarf, with an additional
small correction for our neglect of weak interactions that slightly
decrease \Ye. The difference between the slope given by the detailed
W7-like models and the analytical model is less than 5\%.

Previous investigations (\citealt{hoeflich98}; D01) into the effect of a
varying $X(\Ne)$ do not agree on the mass of \Ni\ produced.
\citet{hoeflich98} found that varying the metallicity from $Z =
0.1\Zsun$ to $Z=10\Zsun$ produced only an $\approx 4\%$ variation in the
\Ni\ mass ejected, in sharp contrast to our results as well the results
of \citet{iwamoto99}, who found that increasing $Z$ from zero to solar
in fast-deflagration models leads to an $\approx 10\%$ decrease in the
\Ni\ mass ejected.  Both of these works relied on post-processing the
thermodynamical trajectories, as we have done.  Recently, D01
investigated the range $10^{-10}\usp\Zsun$ to $0.02\usp\Zsun$.  Of the
one-dimensional calculations just mentioned, only this one accounts for
the effect of \Ye\ on the energy generation rate.  The long-dashed curve
in Fig.~\ref{fig:mni56} shows the D01 results for their 1.5\usp\Msun\
progenitors; other progenitor masses in their survey display the same
trend with $Z$. Our analytical result and post-processed W7-like
models essentially agree with the findings of D01 over the range of
metallicities common to all three calculations.  As is evident in
Fig.~\ref{fig:mni56}, the largest variation in the mass of \Ni\ occurs
at $Z > \Zsun$.  We note that the calculations of D01 evolve a
main-sequence star into the \SNeIa\ progenitor, whereas our calculation
and those of \citet{iwamoto99} start from a given white dwarf
configuration.

As a caveat, we note that our post-processing of the W7 thermodynamic
trajectories is not completely self-consistent.  The reason is that the
temperature and density profiles of the W7 were calculated using the
energy released by burning matter of solar $Z$.  While $Z$ is likely to
influence the flame propagation \citep[via the change in the rate of
energy production;][]{hix96,hix99}, in the mass range under consideration
Si-burning is complete, so our assumption of NSE still holds.  Indeed,
\citet{dominguez00} found that the mass of \Ni\ synthesized was also
rather insensitive to the details of the flame microphysics.  It is also
possible the density, $\rho_{\mathrm{tr}}$, at which a transition from
deflagration to detonation occurs will influence the amount of \Ni\
produced \citep*[][and references therein]{hoeflich95}.  Fits to
observations seem to require, however, that $\rho_{\mathrm{tr}} \approx
10^7\usp\grampercc$ which corresponds to a Lagrangian mass coordinate of
$\sim$ 1.0\usp\Msun, which is exterior to the main \Ni-producing layers.
Further numerical studies should elucidate the source of the
discrepancy. In particular they should determine if other effects, such
as a systematic change in the mass coordinates where most of the \Ni\ is
produced, offsets or enhances the variation in $M(\Ni)$ that we find.
The demonstrated linear dependence between $M(\Ni)$ and $Z$ will make it
easier to untangle competing physical effects in a fully self-consistent
calculation.

\section{Scatter in the Initial Metallicity and the Induced Brightness
  Variations}
\label{sec:scatter-metallicity}

Stellar abundance determinations are discussed in terms of an elemental
abundance relative to iron, \XFe, as a function of the iron to hydrogen
ratio \FeH, primarily because \FeH\ is relatively easy to measure in
stars. The \FeH\ ratio represents a chronometer in that the accumulation
of iron in the ISM increases monotonically with time
\citep*{wheeler89}. Calibration of \FeH\ as a function of time forms the
basis of the age-metallicity relationship.

The $Z$ of local field stars rapidly increased about 10--13\usp\Gyr\ ago
during formation of the Galaxy's disk and then increased much more
gradually over the last $\sim 10\usp\Gyr$ (\citealt{twarog80};
\citealt{edvard93}; \citealt*{feltzing01}).  More importantly for our
purposes, however, is the relatively large scatter in stellar
metallicities, $\Delta\FeH \sim 0.5$, at any given age.
\citet{feltzing01} constructed an age-metallicity diagram for 5828 dwarf
and sub-dwarf stars from the Hipparcos Catalog using evolutionary tracks
to derive ages and Str\"omgren photometry to derive metallicities.  They
concluded that the age-metallicity diagram is well-populated at all
ages, that old, metal-rich stars do exist, and that the scatter in $Z$
at any given age is larger than the observational errors.  Other surveys
of stellar metallicities (\citealt{edvard93}; \citealt{chen00}) are in
good agreement with these trends.

The most abundant elements in the Galaxy after H and He are CNO.  Both
\CFe\ and \NFe\ in halo and disk dwarfs are observed to be roughly solar
and constant (\citealt{laird85}; \citet{carbon87}; \citet{wheeler89}).
The \OFe\ ratio is larger at low metallicities---oxygen being the
dominant element ejected by \SNeII---and then slowly decreases due to
variations in mass and $Z$ (\citealt{gratton85}; \citealt*{peterson90};
\citealt*{timmes95}).  Within these general trends is a relatively large
scatter, $\Delta\CFe \sim \Delta\NFe \sim \Delta\OFe \sim
\unit{0.5}{dex}$, at any given \FeH.

According to the simple analytical relation (eq.~[\ref{eq:x56}]) and the
detailed W7-like models (Fig.~\ref{fig:mni56}), a scatter of a factor of
3 about the mean in the initial metallicity ($\onethird\Zsun < Z <
3\Zsun$) leads to a variation of about 25\% (0.13\usp\Msun) in the mass
of \Ni\ ejected by \SNeIa\ if the \Ne\ and \Fe[56] are uniformly
distributed within the white dwarf.  The minimum peak brightness
variations caused by this variation in \Ni\ mass are $|\Delta
M_V|\approx 0.2$ \citep{pinto01}.  Thus, the amplitude of this effect
cannot account for all of the observed variation in peak luminosity of
local \SNeIa\ (0.5~magnitude in $B$ and $V$).  The observed scatter in
peak brightnesses may be even larger, as Cepheid-based distances to the
host galaxies of peculiar events such as sub-luminous SN~1991bg or
brighter-than-normal SN~1991T haven't been measured yet
(\citealt{saha99}; \citealt*{contardo00}; \citealt{leibundgut00}).
There is evidence for a larger scatter when more distant supernovae are
included \citep{hamuy96, riess98}.  It would take a scatter of about a
factor of seven ($\Delta\FeH\sim \unit{0.8}{dex}$) in the initial
$Z$ to account for a factor of two variation in the \Ni\ mass
and peak luminosity.

\section{Implications and Summary}
\label{sec:implications}

Using the properties of NSE, we find that where weak interactions are
unimportant the mass fraction of \Ni\ produced depends linearly upon the
initial metallicity of the white dwarf progenitor.  This result is
robust: so long as the region that reaches NSE does so on a timescale
over which \Ye\ is nearly constant, then the mass of \Ni\ produced is
largely independent of the detailed physics of the flame front
propagation.  This fact offers a check on sophisticated numerical
calculations and could be exploited to understand better the
disagreements between different codes.

The variation from \onethird\ to 3 times solar metallicity observed in
dwarf stars in the Galactic disk implies that \SNeIa\ should have a
minimum variation of 25\% in the ejected \Ni\ mass.  This has
implications for the brightness variations observed in both near and
distant \SNeIa\ and for galactic chemical evolution.  The most distant
\SNeIa\ observed today have redshifts $z\lesssim 1$.  This corresponds
to a lookback time of about 4--7\usp\Gyr\, depending on the cosmological
model, and implies a mean \FeH\ ratio between -0.1 and -0.3.  There is,
however, still a scatter of $\Delta\FeH\sim 0.5$ at these mean \FeH\
ratios (see the discussion in \S~3).  We therefore expect that the
variation in peak luminosities of \SNeIa\ in spiral galaxies at
$z\lesssim 1$ will show the same minimum variation of $|\Delta
M_V|\approx 0.2$ in peak luminosity as nearby \SNeIa.  This variation is
superposed on other evolutionary effects, such as that from a reduction
in the C/O ratio (\citealt{hoeflich98}; D01).

\lastpagefootnotes
For combustion to NSE with no freeze-out (see \S~2), the \Ni\ mass is
anticorrelated with the mass of \Fe\ and \Ni[58] ejected, and
subluminous \SNeIa\ will tend, therefore, to have larger \FeFe\ ratios
than brighter ones.  Because \SNeIa\ produce $>\onehalf$ of the galactic
iron-peak nuclei\footnote{\SNeIa\ produce $\approx 0.8\usp\Msun$ of
  iron-group nuclei while Type II SNe produce $\approx 0.1\usp\Msun$,
  and the ratio of thermonuclear to core collapse events is about
  0.15--0.27 in the Galaxy \citep{bergh91,cappellaro97}.}, the isotopic
ratios among the iron group in \SNeIa\ ejecta should not exceed the
solar ratios by about a factor of two\citep{wheeler89,iwamoto99}.  There
are several uncertainties with the \FeFe\ ratio.  First, some of the
\Fe\ is produced in the core where weak interactions are important for
determining the final \Ye\ (although this may be alleviated by the
overturning of matter in the core from Rayleigh-Taylor instabilities;
see \citealt{gamezo03} for a recent calculation).  At high densities
these reaction rates are sensitive to the input nuclear physics
(\citealt{brachwitz00}; \citealt*{martinez00}).  Isotopic measurements
of cosmic rays, such as with \emph{Ulysses} (\citealt{connell97};
\citealt{connell01}) and \emph{ACE} \citep{wiedenbeck01} can probe the
evolution of \FeFe\ over the past $\approx 5\usp\Gyr$.

Since most of the \Ni\ created in one-dimensional \SNeIa\ models lies in
the 0.2\usp\Msun--0.8\usp\Msun\ mass shell, our assumption of a uniform
\Ne\ distribution deserves close scrutiny.  There has long been
speculation about sedimentation of \Ne\ and its effect on the cooling of
isolated white dwarfs \citep[][and references
  therein]{hansen02,bildsten01,deloye02}.  If, for example, all of the
\Ne\ from the outermost 0.6\usp\Msun\ were to settle into the shell
between 0.2\usp\Msun\ and 0.8\usp\Msun, then the effective \Ye\ in this
shell would double.  Indeed, \citet{bildsten01} noted that the
production of \Fe\ was an indirect test of the sedimentation of \Ne.
For nearby \SNeIa, which presumably sample a range of progenitor ages,
the variability in \Ni\ production would increase to 50\%, or about
$\Delta M_v\approx 0.3$.  Since it takes about 7\usp\Gyr\ for \Ne\ at
the surface to fall through the outer $\approx 0.4\usp\Msun$ of a
$1.2\usp\Msun$ CO white dwarf (\citealt{bildsten01}), the effect of
sedimentation will be diminished for those \SNeIa\ at $z\approx 1$.

\acknowledgements 

We thank Franziska Brachwitz and Friedel Thielemann for providing the
initial W7 thermodynamic trajectories.  We also thank the referees for
their critical reading, which greatly improved the manuscript. This work
is supported by the Department of Energy under Grant No.~B341495 to the
Center for Astrophysical Thermonuclear Flashes and Grant
No.~DE-FG02-91ER40606 in Nuclear Physics and Astrophysics at the
University of Chicago.


\begin{thebibliography}{99}

\bibitem[Arnett(1982)]{arnett82} 
         Arnett, W.~D.\ 1982, \apj, 253, 785 

\bibitem[Arnett et al.(1971)Arnett, Truran, \& Woosley]{arnett71} 
         Arnett, W.~D., Truran, J.~W., \& Woosley, S.~E.\ 1971, \apj,
         165, 87

\bibitem[van den Bergh \& Tammann(1991)]{bergh91} 
         van den Bergh, S.~\& Tammann, G.~A.\ 1991, \araa, 29, 363 

\bibitem[Bildsten \& Hall(2001)]{bildsten01}
         Bildsten, L. \& Hall, D. 2001, \apjl, 549, L219

\bibitem[Brachwitz et al.(2000)]{brachwitz00} 
         Brachwitz, F.~et al.\ 2000, \apj, 536, 934 

\bibitem[Branch(1998)] {branch98}
         Branch, D.\ 1998, \araa, 36, 17

\bibitem[Cappellaro et al.(1997)]{cappellaro97} 
         Cappellaro, E., Turatto, M., Tsvetkov, D.~Y., Bartunov, O.~S.,
         Pollas, C., Evans, R., \& Hamuy, M.\ 1997, \aap, 322, 431

\bibitem[Carbon et al.(1987)]{carbon87} 
         Carbon, D.~F., Barbuy, B., Kraft, R.~P., Friel, E.~D., 
         \& Suntzeff, N.~B.\ 1987, \pasp, 99, 335 

\bibitem[Chen et al.(2000)]{chen00} 
         Chen, Y.~Q., Nissen, P.~E., Zhao, G., Zhang, H.~W., 
         \& Benoni, T.\ 2000, \aaps, 141, 491 

\bibitem[Clifford \& Tayler(1965)]{clifford65} 
         Clifford, F.~E., \& Tayler, R.~F.\ 1965, \mnras, 129, 104 

\bibitem[Connell(2001)]{connell01}
         Connell, J.~J. 2001, \ssr, 99, 41

\bibitem[Connell \& Simpson(1997)]{connell97}
         Connell, J.~J., \& Simpson, J.~A. 1997, \apjl, 475, L61

\bibitem[Contardo et al.(2000)Contardo, Leibundgut, \& Vacca]{contardo00}
         Contardo, G., Leibundgut, B., \& Vacca, W.~D.\ 2000, \aap, 359, 876 

\bibitem[Deloye \& Bildsten(2002)]{deloye02}
         Deloye, C.~J. \& Bildsten, L. 2002, \apj, 580, 1077

\bibitem[Dom\'inguez \& H\" oflich(2000)]{dominguez00} 
         Dom\'inguez, I.~\& H\" oflich, P.\ 2000, \apj, 528, 854 

\bibitem[Dom\'inguez et al.(2001)Dom\'inguez, H\"oflich, \& Straniero]{dominguez01}
         Dom\'inguez, I., H\"oflich, P., \& Straniero, O.\ 2001, \apj,
         557, 279 (D01)

\bibitem[Edvardsson et al.(1993)]{edvard93} 
         Edvardsson, B., Andersen, J., Gustafsson, B., Lambert, D.~L., 
         Nissen, P.~E., \& Tomkin, J.\ 1993, \aap, 275, 101 

\bibitem[{{Feltzing} {et~al.}(2001){Feltzing}, {Holmberg}, \&
    {Hurley}}]{feltzing01} 
         Feltzing, S., Holmberg, J., \& Hurley, J.~R.\ 2001, \aap, 377, 911 

\bibitem[Fillipenko(1997)]{filippenko97}
         Fillipenko, A.~V.\ 1997, \araa, 35, 309 

\bibitem[Gamezo et al.(2003)]{gamezo03}
         Gamezo, V.~N., Khokhlov, A.~M., Oran, E.~S., Chtchelkanova,
         A.~Y., \& Rosenberg, R.~O.\ 2003, Science, 299, 77 

\bibitem[Gibson et al.(2000)]{gibson00}
         Gibson, B.~K. et al. \ 2000, \apj, 529, 723 

\bibitem[Gratton(1985)]{gratton85} 
         Gratton, R.~G.\ 1985, \aap, 148, 105 

\bibitem[Hamuy et al.(1996)]{hamuy96}
         Hamuy, M., Phillips, M.~M., Suntzeff, N.~B., Schommer, R.~A.,
         Maza, J., Smith, R.~C., Lira, P., \& Aviles, R.\ 1996, \aj, 112,
         2438

\bibitem[Hansen et al.(2002)]{hansen02} 
         Hansen, B.~M.~S.~et al.\ 2002, \apjl, 574, L155 

\bibitem[Hix \& Thielemann(1996)]{hix96}
         Hix, W.~R., \& Thielemann, F.-K. 1996, \apj, 460, 869

\bibitem[Hix \& Thielemann(1999)]{hix99}
         Hix, W.~R., \& Thielemann, F.-K. 1999, \apj, 511, 862

\bibitem[H\"oflich et al.(1995)H\"oflich, Khokhlov, \& Wheeler]{hoeflich95}
         H\"oflich, P., Khokhlov, A. M., \& Wheeler, J. C. 1995, 
         \apj, 444, 831

\bibitem[H\"oflich et al.(1998)H\"oflich, Wheeler, \& Thielemann]{hoeflich98} 
         H\"oflich, P., Wheeler, J.~C., \& Thielemann, F.~K.\ 1998,
         \apj, 495, 617

\bibitem[H\"oflich et al.(2000)]{hoeflich00} 
         H\"oflich, P., Nomoto, K., Umeda, H., 
         \& Wheeler, J.~C.\ 2000, \apj, 528, 590

\bibitem[Iwamoto et al.(1999)]{iwamoto99} 
         Iwamoto, K., Brachwitz, F., Nomoto, K., Kishimoto, N., 
         Umeda, H., Hix, W.~R., \& Thielemann, F.\ 
         1999, \apjs, 125, 439 

\bibitem[Laird(1985)]{laird85} 
         Laird, J.~B.\ 1985, \apj, 289, 556 

\bibitem[Leibundgut(2000)]{leibundgut00}
         Leibundgut, B.\ 2000, A\&A rev., 10, 179

\bibitem[Leibundgut(2001)]{leibundgut01}
         Leibundgut, B.\ 2001, \araa, 39,67

\bibitem[Mart{\'{\i}}nez-Pinedo et al.(2000)Mart{\'{\i}}nez-Pinedo,
  Langanke, \& Dean]{martinez00} 
         Mart{\'{\i}}nez-Pinedo, G., Langanke, K., \& Dean, D.~J.
         \ 2000, \apjs, 126, 493 

\bibitem[Nomoto et al.(1984)Nomoto, Thielemann, \& Yokoi]{nomoto84} 
         Nomoto, K., Thielemann, F.-K., \& Yokoi, K.\ 1984, \apj, 286,
         644

\bibitem[Perlmutter et al.(1999)]{perlmutter99}
         Perlmutter, S., et al. \ 1999, \apj, 517, 565

\bibitem[Peterson et al.(1990)Peterson, Kurucz, \& Carney]{peterson90} 
         Peterson, R.~C., Kurucz, R.~L., \& Carney, B.~W.\ 1990, \apj,
         350, 173

\bibitem[Pinto \& Eastman(2000)]{pinto00}
         Pinto, P.~A. \& Eastman, R.~G. 2000, \apj, 530, 744

\bibitem[Pinto \& Eastman(2001)]{pinto01}
         Pinto, P.~A. \& Eastman, R.~G. 2001, New Astronomy, 6, 307

\bibitem[Phillips et al.(1999)]{phillips99}
         Phillips, M.~M., Lira, P., Suntzeff, N.~B., Schommer, R.~A., 
         Hamuy, M., Maza, J., \aj, 118. 1766

\bibitem[Riess et al.(1998)]{riess98}
         Riess, A.~G. et al. \ 1998, \aj, 116, 1009

\bibitem[Saha et al.(1999)]{saha99} 
         Saha, A., Sandage, A., Tammann, G.~A., Labhardt, L., 
         Macchetto, F.~D., \& Panagia, N.\ 1999, \apj, 522, 802 

\bibitem[Thielemann et al.(1986)Thielemann, Nomoto, \&Yokoi]{thielemann86} 
         Thielemann, F.-K., Nomoto, K., \& Yokoi, K.\ 1986, \aap, 158, 17 

\bibitem[Timmes et al.(1995)Timmes, Woosley, \& Weaver]{timmes95} 
         Timmes, F.~X., Woosley, S.~E., \& Weaver, T.~A. \ 1995, \apjs,
         98, 617

\bibitem[Truran et al.(1967)Truran, Arnett, \& Cameron]{truran67} 
         Truran, J.~W., Arnett, D., \& Cameron, A.~G.~W. \ 1967,
         Canad.~J.~Phys., 45, 2315

\bibitem[Twarog(1980)]{twarog80} 
         Twarog, B.~A.\ 1980, \apj, 242, 242 

\bibitem[Umeda et al.(1999a)]{umeda99a}
         Umeda, H., Nomoto, K., Kobayashi, C., Hachisu, I., \& Kato,
         M.\ 1999, \apjl, 522, L43

\bibitem[Umeda et al.(1999b)]{umeda99b}
         Umeda, H., Nomoto, K., Yamaoka, H., \& Wanajo, S.\ 1999, \apj,
         513, 861

\bibitem[Wheeler et al.(1989)Wheeler, Sneden, \& Truran]{wheeler89}
         Wheeler, J.~C., Sneden, C., \& Truran,~J. W. \ 1989, \araa, 27,
         279

\bibitem[Wiedenbeck et al.(2001)]{wiedenbeck01}
         Wiedenbeck, M. E., et al.\ 2001, \ssr, 99, 15

\end{thebibliography}
\end{document}